\documentclass[aps,preprint]{revtex4}
\usepackage{graphics}

\begin{document}

\preprint{Version for CUP 04-07-03}
\title[Conceptual tensions between QM and GR]{Conceptual tensions between
quantum mechanics and general relativity: Are there experimental
consequences?}
\author{Raymond Y. Chiao}
\affiliation{Department of Physics\\
University of California\\
Berkeley, CA 94720-7300\\
U. S. A.\\
(E-mail: chiao@physics.berkeley.edu)}
\keywords{General Relativity, Quantum Mechanics}
\pacs{PACS number}

\begin{abstract}
One\ of the conceptual tensions between quantum mechanics (QM) and general
relativity (GR) arises from the clash between the \textit{spatial
nonseparability} of entangled states in QM, and the complete \textit{spatial
separability} of all physical systems in GR, i.e., between the $nonlocality$
implied by the superposition principle, and the $locality$ implied by the
equivalence principle. \ Possible experimental consequences of this
conceptual tension will be discussed for macroscopically entangled, coherent
quantum fluids, such as superconductors, superfluids, atomic Bose-Einstein
condensates, and quantum Hall fluids, interacting with tidal and
gravitational radiation fields. ~A minimal-coupling rule, which arises from
the electron spin coupled to curved spacetime, leads to an interaction
between electromagnetic (EM) and gravitational (GR) radiation fields
mediated by a quantum Hall fluid. ~This suggests the possibility of a
quantum transducer action, in which EM waves are convertible to GR waves,
and vice versa.
\end{abstract}

\volumeyear{year}
\volumenumber{number}
\issuenumber{number}
\eid{identifier}
\date[Date of final version for Cambridge University Press: ]{April 7, 2003}
\maketitle

\address{Department of Physics\\
University of California\\
Berkeley, CA 94720-7300\\
U. S. A.\\
(E-mail: chiao@physics.berkeley.edu)}

\section*{Introduction}

\begin{quotation}
\bigskip \emph{``Mercy and Truth are met together; Righteousness and Peace
have kissed each other.''} \ (Psalm 85:10)
\end{quotation}

In this Festschrift Volume in honor of John Archibald Wheeler, I would like
to take a fresh look at the intersection between two fields to which he
devoted much of his research life: general relativity (GR) and quantum
mechanics (QM). As evidence of his keen interest in these two subjects, I
would cite two examples from my own experience. When I was an undergraduate
at Princeton University during the years from 1957 to 1961, he was my
adviser. One of his duties was to assign me topics for my junior paper and
for my senior thesis. For my junior paper, I was assigned the topic: Compare
the complementarity and the uncertainty principles of quantum mechanics:
Which is more fundamental? For my senior thesis, I was assigned the topic:
How to quantize general relativity? As Wheeler taught me, more than half of
science is devoted to the asking of the right question, while often less
than half is devoted to the obtaining of the correct answer, but not always!
\ 

In the same spirit, I would like to offer up here some questions concerning
conceptual tensions between GR and QM, which hopefully can be answered in
the course of time by experiments, with a view towards probing the tension
between the concepts of $locality$ in GR and $nonlocality$ in QM. \ I hope
that it would be appropriate and permissible to ask some questions here
concerning this tension. \ It is not the purpose of this Chapter to present
demonstrated results, but to suggest heuristically some interesting avenues
of research which might lead to future experimental discoveries. \ 

One question that naturally arises at the border between GR and QM is the
following: Are there novel experimental or observational ways of studying
quantized fields coupled to curved spacetime? ~This question has already
arisen in the context of the vacuum embedded in curved spacetime \cite%
{DaviesBook}, but I would like to extend this to possible experimental
studies of the ground state of a nonrelativistic quantum many-body system
with off-diagonal long-range order, i.e., a ``quantum fluid,'' viewed as a
quantized field, coupled to curved spacetime. As we shall see, this will
naturally lead to the further question: Are there $quantum$ methods to
detect gravitational radiation other than the $classical$ ones presently
being used in the Weber bar and LIGO (i.e., the ``Laser Interferometer
Gravitational Wave Observatory'') \cite{MTW}\cite{Tyson}\cite{Taylor}?

As I see it, the three main pillars of physics at the beginning of the 21st
century are quantum mechanics, relativity, and statistical mechanics, which
correspond to Einstein's three papers of 1905. There exist conceptual
tensions at the intersections of these three fields of physics (see Figure
1). It seems worthwhile re-examining these tensions, since they may entail
important experimental consequences. In this introduction, I shall only
briefly mention three conceptual tensions between these three fields: $%
locality$ versus $nonlocality$ of physical systems, $objectivity$ versus $%
subjectivity$ of probabilities in quantum and statistical mechanics (the
problem of the nature of information), and $reversibility$ versus $%
irreversibility$ of time (the problem of the arrows of time). \ Others in
this Volume will discuss the second and\ the third of these tensions in
detail. \ I shall limit myself to a discussion of the first conceptual
tension concerning locality versus nonlocality, mainly in the context of GR
and QM. \ (However, in my Solvay lecture \cite{ChiaoSolvay}, I have
discussed the other two tensions in more detail. See also my Rome lecture %
\cite{ChiaoRome} for a discussion of three different kinds of quantum
nonlocalities).

Why examine conceptual tensions? \ A brief answer is that they often lead to
new experimental discoveries. \ It suffices to give just one example from
late 19th and early 20th century physics: the clash between the venerable
concepts of $continuity$ and $discreteness$. \ The concept of continuity,
which goes back to the Greek philosopher Heraclitus (``everything flows''),
clashed with the concept of discreteness, which goes back to Democritus
(``everything is composed of atoms''). \ Eventually, Heraclitus's concept of
continuity, or more specifically that of the $continuum$, was embodied in
the idea of $field$ in the classical field theory associated with Maxwell's
equations. \ The atomic hypothesis of Democritus was eventually embodied in
the kinetic theory of gases in statistical mechanics. \ 

Conceptual tensions, or what Wheeler calls the ``clash of ideas,'' need not
lead to a complete victory of one conflicting idea over the other, so as to
eliminate the opposing idea completely, as seemed to be the case in the 19th
century, when Newton's idea of ``corpuscles of light'' was apparently
completely eliminated in favor of the wave theory of light. \ Rather, there
may result a reconciliation of the two conflicting ideas, which\ then often
leads to many fruitful experimental consequences. \ 

Experiments on blackbody radiation in the 19th century were exploring the
intersection, or borderline, between Maxwell's theory of electromagnetism
and statistical mechanics, where the conceptual tension between continuity
and discreteness was most acute, and eventually led to the discovery of
quantum mechanics through the work of Planck. The concept of $discreteness$
metamorphosed into the concept of the $quantum$. \ This led\ in turn to the
concept of \textit{discontinuity} embodied in Bohr's \textit{quantum jump}
hypothesis, which was necessitated by the indivisibility of the quantum.
Many experiments, such as Millikan's measurements of $h/e$, were in turn
motivated by Einstein's heuristic theory of the photoelectric effect based
on the ``light quantum'' hypothesis. \ Newton's idea of ``corpuscles of
light'' metamorphosed into the concept of the $photon$. \ This is a striking
example showing how that many fruitful experimental consequences can come
out of one particular conceptual tension.

Within a broader cultural context, there have been many acute conceptual
tensions between science and faith, which have lasted over many centuries. \
Perhaps the above examples of the fruitfulness of the resolution of
conceptual tensions within physics itself may serve as a parable concerning
the possibility of a peaceful reconciliation of these great cultural
tensions, which may eventually lead to the further growth of both science
and faith. \ Hence we should not shy away from conceptual tensions, but
rather explore them with an honest, bold, and open spirit.

\section{Three conceptual tensions between quantum mechanics and general
relativity}

Here I shall focus my attention on some specific conceptual tensions at the
intersection between QM and GR. \ A commonly held viewpoint within the
physics community today is that the only place where conceptual tensions
between these two fields can arise is at the microscopic Planck length scale
($1.6\times 10^{-33}$ cm), where quantum fluctuations of spacetime
(``quantum foam'') occur. \ Hence manifestations of these tensions would be
expected to occur only in conjunction with extremely high-energy phenomena,
accessible presumably only in astrophysical settings, such as the early Big
Bang. \ \qquad

However, I believe that this point of view is too narrow. \ There exist
other\ conceptual tensions at macroscopic, non-Planckian\ distance scales ($%
\gg 1.6\times 10^{-33}$ cm), which should be accessible in low-energy
laboratory experiments involving macroscopic QM phenomena. \ It should be
kept in mind that QM not only describes $microscopic$ phenomena, but also $%
macroscopic$ phenomena, such as superconductivity. \ Specifically, I would
like to point out the following\ three conceptual tensions:

\begin{description}
\item (1) The \textit{spatial nonseparability} of physical systems due to
entangled states in QM, versus the complete \textit{spatial separability} of
all physical systems in GR.

\item (2) The \textit{equivalence principle} of GR, versus the \textit{%
uncertainty principle} of QM.

\item (3) The \textit{mixed state} (e.g., of an entangled bipartite system,
one part of which falls into a black hole; the other of which flies off to
infinity) in GR, versus the \textit{pure state} of such a system in QM.
\end{description}

Conceptual tension (3) concerns the problem of the natures of information
and entropy in QM and GR. \ Again, since others will discuss this tension in
detail in this Volume, I shall limit myself only to a discussion of the
first two of these tensions. \ 

These conceptual tensions originate from the \textit{superposition principle}
of QM, which finds its most dramatic expression in the \textit{entangled
state} of two or more spatially separated particles of a single physical
system, which in turn leads to Einstein-Podolsky-Rosen\ (EPR) effects. \ It
should be emphasized here that it is necessary to consider \textit{two or
more} particles for observing EPR phenomena, since only then does the 
\textit{configuration space} of these particles no longer coincide with that
of ordinary spacetime. \ For example, consider the entangled state of two
spin 1/2 particles in a ``singlet'' state initially prepared in the total
spin zero state 
\begin{equation}
\left| S=0\right\rangle =\frac{1}{\sqrt{2}}\left\{ \left| \uparrow
\right\rangle _{1}\left| \downarrow \right\rangle _{2}-\left| \downarrow
\right\rangle _{1}\left| \uparrow \right\rangle _{2}\right\} \text{ ,}
\label{Bohm singlet}
\end{equation}%
in which the two particles in a spontaneous decay process fly arbitrarily
far away from each other into two space-like separated regions of spacetime,
where measurements on spin by means of two Stern-Gerlach apparati are
performed separately on these two particles.

As a result of the quantum entanglement arising from the $superposition$ of $%
product$ states, such as in the above singlet state suggested by Bohm in
connection with the EPR ``paradox,'' it is in general impossible to
factorize this state into products of probability amplitudes. \ Hence it is
impossible to factorize the $joint$ probabilities in the measurements of
spin of this two-particle system. \ This mathematical $nonfactorizability$
implies a physical $nonseparability$ of the system, and leads to
instantaneous, space-like correlations-at-a-distance in the joint
measurements of the properties (e.g., spin) of discrete events, such as in
the coincidence detection of ``clicks'' in Geiger counters placed behind the
two distant Stern-Gerlach apparati. \ Bell's inequalities place an upper
limit the amount of angular correlations possible for these two-particle
decays, based on the $independence$ (and hence $factorizability$) of the
joint probabilities of spatially separated measurements in any local
realistic theories, such as those envisioned by Einstein. \ 

Violations of Bell's inequalities have been extensively experimentally
demonstrated \cite{Gisin2002}. \ Therefore these observations cannot be
explained on the basis of any \textit{local realistic} world view; however,
they were predicted by QM. \ If we assume a $realistic$ world view, i.e.,
that the ``clicks'' of the Geiger counters really happened, then we must
conclude that we have observed $nonlocal$ features of the world. \ Therefore
a fundamental \textit{spatial nonseparability} of physical systems has been
revealed by these Bell-inequalities-violating EPR experiments \cite{Chiao
and Garrison}. \ It should be emphasized that the observed space-like EPR
correlations occur on macroscopic, non-Planckian distance scales, where the
conceptual tension (1) between QM and GR becomes most acute.

Although some of these same issues arise in the conceptual tensions between
quantum mechanics and $special$ relativity, there are new issues which crop
up due to the long-range nature of the gravitational force, which are absent
in special relativity, but present in $general$ relativity. The problem of
quantum fields in $curved$ spacetime can be more interesting than in $flat$
spacetime.

Gravity is a \textit{long-range} force. \ It is therefore\ natural\ to
expect that experimental consequences of conceptual tension (1)\ should
manifest themselves most dramatically in the\ interaction of macroscopically
coherent quantum matter, which exhibit \textit{long-range} EPR correlations,
with \textit{long-range} gravitational fields. \ In particular, the question
naturally arises: How do entangled states, such as the above singlet state,
interact with tidal fields, such as those in gravitational radiation? \
Stated more generally: How do quantum many-body systems with entangled
ground states possessing off-diagonal long-range order couple to curved
spacetime? (``Off-diagonal long-range order'' (ODLRO) means that the
off-diagonal elements of the reduced density matrix in a coordinate space
representation of the system are nonvanishing and possess long-range order,
i.e., macroscopic quantum phase coherence.) It is therefore natural to look
to the realm of\ \textit{macroscopic} phenomena associated with quantum
fluids, rather than phenomena at \textit{microscopic}, Planck length scales,
in our search for these experimental consequences.

Already a decade or so before Bell's ground-breaking work on his inequality,
Einstein himself was clearly worried by the radical, spatial nonseparability
of physical systems in quantum mechanics. \ Einstein wrote \cite%
{Einstein-Born}:

\begin{quotation}
``Let us consider a physical system S$_{12}$, which consists of two
part-systems S$_{1}$ and S$_{2}$. \ These two part-systems may have been in
a state of mutual physical interaction at an earlier time. \ We are,
however, considering them at a time when this interaction is at an end. \
Let the entire system be completely described in the quantum mechanical
sense by a $\mathrm{\psi }$-function $\mathrm{\psi }_{12}$ of the
coordinates $\mathrm{q}_{1}$,... and $\mathrm{q}_{2}$,... of the two
part-systems ($\mathrm{\psi }_{12}$ cannot be represented as a product of
the form $\mathrm{\psi }_{1}\mathrm{\psi }_{2}$ but only as a sum of such
products [\textit{i.e., as an entangled state}]). \ At time\textrm{\ }t%
\textrm{\ }let the two part-systems be separated from each other in space,
in such a way that $\mathrm{\psi }_{12}$ only differs from zero when $%
\mathrm{q}_{1}$,... belong to a limited part R$_{1}$ of space and \ $\mathrm{%
q}_{2}$,... belong to a part R$_{2}$ separated from R$_{1}$. \ \ \ . . .

``There seems to me no doubt that those physicists who regard the
descriptive methods of quantum mechanics as definitive in principle would
react to this line of thought in the following way: they would drop the
requirement for \textit{the independent existence of the physical reality
present in different parts of space}; they would be justified in pointing
out that the quantum theory nowhere makes explicit use of this
requirement.'' [\textit{Italics mine.}]
\end{quotation}

This radical, \textit{spatial nonseparability} of a physical system
consisting of two or more entangled particles in QM, which seems to
undermine the very possibility of the concept of \textit{field} in physics,
is in an obvious conceptual tension with the\ complete spatial separability
of any physical system into\ its separate parts in GR, which is a \textit{%
local realistic} field theory. \ 

However, I should hasten to add immediately that the battle-tested concept
of \textit{field} has of course been extremely fruitful not only at the
classical but also at the quantum level. \ Relativistic quantum field
theories have been very well validated, at least in an approximate,
correspondence-principle sense in which spacetime itself is treated
classically, i.e., as being describable by a rigidly flat, Minkowskian
metric, which has no\ possibility of any quantum dynamics. \ There have been
tremendous successes of quantum electrodynamics and electroweak gauge field
theory (and, to a lesser extent, quantum chromodynamics) in passing all
known high-energy experimental tests. \ Thus the conceptual tension between $%
continuity$ (used in the concept of the spacetime $continuum$) and $%
discreteness$ (used in the concept of \textit{quantized excitations} of a
field in classical spacetime) seems to have been successfully reconciled in
these relativistic quantum field theories. \ Nevertheless, the problem of a
satisfactory relativistic treatment of quantum measurement within these
theories remains an open one \cite{Peres}.

\section{Is there any difference between the response of classical and
quantum fluids to tidal gravitational fields?}

Motivated by the above discussion, a more specific question arises: Is there
any difference between classical and quantum matter when it is embedded in
curved spacetime, for instance, in the $linear$ response to the
gravitational tidal field of the Earth of a $classical$ liquid drop, as
compared to that of a $quantum$ one, such as a liquid drop of superfluid
helium? \ In order to answer this question, consider a gedanken experiment
to observe the shape of a freely floating liquid drop placed at the center
of the Space Station sketched in Figure 2.

At first glance, the answer to this question would seem to be ``no,'' since
the equivalence principle would seem to imply that all freely falling
bodies, whether classical or quantum, must respond to gravitation, e.g.,
Earth's gravity, in a \emph{mass-independent}, or more generally, in a \emph{%
composition-independent} way. \ Thus whether the internal dynamics of the
particles composing the liquid drop obeys classical mechanics or quantum
mechanics would seem to make no difference in the response of this body to
gravity. \ Just as in the case of the response of the tides of the Earth's
oceans to the Moon's gravity, the shape of the surface of a liquid of any
mass or composition would be determined by the equipotential surfaces of the
total gravitational field, and should be independent of the mass or
composition of the liquid, provided that the fluid particles can move $%
freely $ inside the fluid, and provided that the surface tension of the
liquid can be neglected.

However, one must carefully distinguish between the response of the \textit{%
center of mass} of the liquid drop inside the Space Station to Earth's
gravity, and the response of the \textit{relative motions} of particles
within the drop to Earth's tidal gravitational field. \ Whereas the former
clearly obeys the mass- and composition-independence of the equivalence
principle, one must examine the latter with more care. \ First, one must
define what one means by ``classical'' and ``quantum'' bodies. \ By a
``classical body,'' we shall mean here a body whose particles have undergone 
$decoherence$ in the sense of Zurek \cite{Zurek}, so that no macroscopic,
Schr\"{o}dinger-cat-like states for~widely spatially separated subsystems
(i.e., the fluid elements inside the classical liquid drop) can survive the
rapid decoherence arising from the environment. This is true for the vast
majority of bodies typically encountered in the laboratory. \ It is the
rapid decoherence of the spatially separated subsystems of a classical body
that makes the \textit{spatial separability} of a system into its parts, and
hence $locality$, a valid concept.

Nevertheless, there exist exceptions. \ For example, a macroscopically
coherent quantum system, e.g., a quantum fluid such as the electron pairs
inside a superconductor, usually possesses an energy gap which separates the
ground state of the system from all possible excited states of the system. \
Cooper pairs of electrons in a Bardeen, Cooper, and Schrieffer (BCS) ground
state are in the entangled spin singlet states given by Eq.(\ref{Bohm
singlet}). At sufficiently low temperatures, such a quantum fluid develops a
macroscopic quantum coherence, as is manifested by a macroscopic quantum
phase which becomes well defined at each point inside the fluid. The
resulting macroscopic wavefunction must remain single valued, in spite of
small perturbations, such as those due to weak external fields.

The energy gap, such the BCS gap, protects spatially separated, but
entangled, particles within the body, such as the electrons which members of
Cooper pairs inside a superconductor, against decoherence. \ Therefore,
these quantum fluids are \textit{protectively entangled}, in the sense that
the existence of some sort of energy gap separates the nondegenerate ground
state of the system from all excited states, and hence prevents any rapid
decoherence due to the environment. Under these circumstances, the
macroscopically entangled ground state of a quantum fluid, becomes a
meaningful $global$ concept, and the notion of $nonlocality$, that is, the 
\textit{spatial nonseparability} of a system into its parts, enters in an
intrinsic way into the problem of the interaction of matter with
gravitational fields.

For example, imagine a liquid drop consisting of superfluid helium at zero
Kelvin, which is in a pure quantum state, floating at the center of the
Space Station, as pictured in Figure 3. Although the microscopic many-body
problem for this superfluid has not been completely solved, there exist a
successful macroscopic, phenomenological description based on the
Gross-Pitaevskii equation ~%
\begin{equation}
-\frac{\hbar ^{2}}{2m}\nabla ^{2}\Psi +V(x,y,z)\Psi +\beta \left| \Psi
\right| ^{2}\Psi =-\alpha \Psi \text{ ,}
\end{equation}%
where $\Psi $ is the macroscopic complex order parameter, and the potential $%
V(x,y,z)$ describes Earth's gravity (including its tidal gravitational
potential, but neglecting for the moment the frame-dragging term coupled to
superfluid currents), along with the surface tension effects which enters
into the determination of the free boundary of the liquid drop. Macroscopic
quantum entanglement is contained in the nonlinear term $\beta \left| \Psi
\right| ^{2}\Psi $, which arises microscopically from atom-atom $S$-wave
scattering events, just as in the case of the recently observed atomic
Bose-Einstein condensates (BECs). (The parameter $\beta $ is directly
proportional to the $S$-wave scattering length $a$; the interaction between
two atoms in a individual scattering event entangles the two scattering
atoms together, so that a measurement of the momentum of one atom
immediately determines the momentum of the other atom which participated in
the scattering event.) ~As in the case of the BECs, where this equation has
been successfully applied to predict many observed phenomena, the physical
meaning of $\Psi $ is that it is the condensate wavefunction.

There should exist near the inside surface~of the superfluid liquid drop,
closed trajectories for helium atom wave packets propagating at grazing
incidence, which, in the correspondence-principle limit, should lead to the
atomic analog of the ``whispering gallery modes'' of light, such as those
observed inside microspheres immersed in superfluid helium \cite{Braginsky}%
\cite{Haroche}. In the case of light, these modes can possess extremely high 
$Q$s (of the order of $10^{9}$), so that the quadrupolar distortion from a
spherical shape due to tidal forces can thereby be very sensitively measured
optically (the degeneracy of these modes has been observed to be split by
nontidal quadrupolar distortions \cite{Chang}). The atomic wave packets
propagating at grazing incidence near the surface are actually those of
individual helium atoms dressed by the collective excitations of the
superfluid, such as phonons, rotons, and ripplons \cite{CriticalAngle}.
Application of the Bohr-Sommerfeld quantization rule to the closed
trajectories which correspond to the whispering gallery modes for atoms
should lead to a $quantization$ of the sizes and shapes of the superfluid
drop. For a classical liquid drop, no such quantization occurs because of
the decoherence of an atom after it has propagated around these large,
polygonal closed trajectories. Hence there should exist a $difference$
between classical and quantum matter in their respective responses to
gravitational tidal fields. At a fundamental level, this difference arises
from the quantum phase shift which is observable in the shift of the
interference fringe pattern that results from an atom travelling coherently
along two nearby, but intersecting, geodesics in the presence of spacetime
curvature \cite{GRGessay}.

Another difference between a classical and a quantum liquid drop is the
possibility of the presence of $quantized$ vortices in the latter, along
with their associated persistent, macroscopic current flows. These quantum
flows possess quantized vorticities of $\pm h/m$, where $m$ is the mass of
the superfluid atom. The question naturally arises: How do two such vortices
placed symmetrically around the center of mass of a superfluid liquid drop
react to the presence of tidal forces associated with gravitational
radiation? I suspect that these vortices will move \emph{at right angles} in
response to these forces in accordance with the Magnus force law, which is a
Lorentz-like force law for vortex motion in superfluids. The
perpendicularity of this kind of motion is manifestly different from that a
test particle of a classical ``perfect'' fluid.

Such $differences$ in the linear response between classical and quantum
matter in the induced quadrupole moment $\Delta Q_{ij}$ of the liquid drop
can be characterized by a linear equation relating $\Delta Q_{ij}$ to the
metric deviations from flat spacetime $h_{kl}$ by means of a
phenomenological susceptibility tensor $\left. \Delta \chi _{ij}\right. ^{kl}
$, viz., 
\begin{equation}
\Delta Q_{ij}=\left. \Delta \chi _{ij}\right. ^{kl}h_{kl}\text{ ,}
\end{equation}%
where $i,j,k,l$ are spatial indices (repeated indices are summed). The
susceptibility tensor $\left. \Delta \chi _{ij}\right. ^{kl}$ should in
principle be calculable from the many-body current-current correlation
function in the linear-response theory of superfluid helium \cite{Forster}.

Here, however, I shall limit myself only to some general remarks concerning $%
\left. \Delta \chi _{ij}\right. ^{kl}$ based on the Kramers-Kronig
relations. Since the response of the liquid drop to weak tidal gravitational
fields is $linear$ and $causal$, it follows that%
\begin{equation}
\mathrm{Re}\left. \Delta \chi _{ij}\right. ^{kl}(\omega )=\frac{1}{\pi }%
P\int_{-\infty }^{\infty }d\omega ^{\prime }\frac{\mathrm{Im}\left. \Delta
\chi _{ij}\right. ^{kl}(\omega ^{\prime })}{\omega ^{\prime }-\omega }
\end{equation}%
\begin{equation}
\mathrm{Im}\left. \Delta \chi _{ij}\right. ^{kl}(\omega )=-\frac{1}{\pi }%
P\int_{-\infty }^{\infty }d\omega ^{\prime }\frac{\mathrm{Re}\left. \Delta
\chi _{ij}\right. ^{kl}(\omega ^{\prime })}{\omega ^{\prime }-\omega }\text{
,}
\end{equation}%
where $P$ denotes Cauchy's Principal Value. From the first of these
relations, there follows the zero-frequency sum rule%
\begin{equation}
\mathrm{Re}\left. \Delta \chi _{ij}\right. ^{kl}(\omega \rightarrow 0)=\frac{%
2}{\pi }\int_{0}^{\infty }d\omega ^{\prime }\frac{\mathrm{Im}\left. \Delta
\chi _{ij}\right. ^{kl}(\omega ^{\prime })}{\omega ^{\prime }}\text{ .}
\end{equation}%
This equation tells us that $if$ there should exist a difference in the
linear response between classical and quantum matter to tidal fields at DC
(i.e., $\omega \rightarrow 0$) in the quadrupolar shape of the liquid drop, $%
then$ there must also exist a difference in the rate of absorption or
emission of gravitational radiation due to the imaginary part of the
susceptibility $\mathrm{Im}\left. \Delta \chi _{ij}\right. ^{kl}(\omega
^{\prime })$ between classical and quantum matter. The purpose here is not
to calculate how big this difference is, but merely to point out that such a
difference exists. The above considerations also apply equally well to an
atomic BEC, indeed, to any quantum fluid, in its linear response to tidal
fields.

\section{Quantum fluids versus perfect fluids \ }

At this point, I would like to return to the more general question: Where to
look for experimental consequences of conceptual tension (1)? ~The above
discussion suggests the following answer: Look at \textit{macroscopically
entangled}, and thus \textit{radically delocalized}, quantum states
encountered, for example, in superconductors, superfluids, atomic BECs, and
quantum Hall fluids, i.e., in what I shall henceforth call ``quantum
fluids.'' \ Again it should be stressed that since gravity is a \textit{%
long-range} force, it should be possible to perform \textit{low-energy}
experiments to probe the interaction between gravity and these kinds of
quantum matter on large, non-Planckian distance scales, without the
necessity of performing high-energy experiments, as is required for probing
the short-range weak and strong forces on very short distance scales. \ The
quantum many-body problem, even in its nonrelativistic limit, may lead to
nontrivial interactions with weak, long-range gravitational fields, as the
above example suggests. \ One is thereby\ strongly motivated to study the
interaction of these quantum fluids with weak gravity, in particular, with
gravitational radiation.

One manifestation of this conceptual tension is that the way one views a
quantum fluid in QM is conceptually radically different from the way that
one views a perfect fluid in GR, where only the $local$ properties of the
fluid, which can conceptually always be spatially separated into
independent, infinitesimal fluid elements, are to be considered. \ For
example, interstellar dust particles can be thought of as being a perfect
fluid in GR, provided that we can neglect all interactions between such
particles \cite{Colin}. \ At a fundamental level, the spatial separability
of the perfect fluid in GR arises from the rapid decoherence of quantum
superposition states (i.e., Schr\"{o}dinger cat-like states) of various
interstellar dust particles at widely separated spatial positions within a
dust cloud, due to interactions with the environment. \ Hence the notion of $%
locality$ is valid here. The response of these dust particles in the
resulting $classical$ many-body system to a gravitational wave passing over
it, is characterized by the local, classical, free-fall motion of each
individual dust particle.

In contrast to the classical case, due to their radical delocalization,
particles in a macroscopically coherent quantum many-body system, i.e., a
quantum fluid, are entangled with each other in such a way that there arises
an unusual ``quantum rigidity'' of the system, closely associated with what
London called ``the rigidity of the macroscopic wavefunction'' \cite{London}%
. \ One example of such a rigid quantum fluid is the ``incompressible
quantum fluid'' in both the integer and the fractional quantum Hall effects %
\cite{Laughlin}. \ This rigidity arises from the fact that there exists an
energy gap (for example, the quantum Hall gap) which separates the ground
state from all the low-lying excitations of the system. This gap, as pointed
out above, also serves to protect the quantum entanglement present in the
ground state from decoherence due to the environment, provided that the
temperature of these quantum systems is sufficiently low. \ Thus these
quantum fluids exhibit a kind of ``gap-protected quantum entanglement.''
~Furthermore, the gap leads to an evolution in accordance with the quantum
adiabatic theorem: The system stays adiabatically in a rigidly unaltered
ground state, which leads in first-order perturbation theory to quantum
diamagnetic effects. Examples of consequences of this ``rigidity of the
wavefunction'' are the Meissner effect in the case of superconductors, in
which the magnetic field is expelled from their interiors, and the
Chern-Simons effect in the quantum Hall fluid, in which the photon acquires
a mass inside the fluid.

\section{Spontaneous symmetry breaking, off-diagonal long-range order, and
superluminality}

The unusual states of matter in these quantum fluids usually possess \textit{%
spontaneous symmetry breaking}, in which the ground state, or the ``vacuum''
state, of the quantum many-body system breaks the symmetry present in the
free energy of the system. \ The physical vacuum, which is in an
intrinsically nonlocal ground state of relativistic quantum field theories,
possesses certain similarities to the ground state of a superconductor, for
example. \ Weinberg has argued that in superconductivity, the spontaneous
symmetry breaking process results in a broken $gauge$ invariance \cite%
{Weinberg(Nambu)}, an idea which traces back to the early work of Nambu \cite%
{Nambu}. \ 

The Meissner effect in a superconductor is closely analogous to the Higgs
mechanism of high-energy physics, in which the physical vacuum also
spontaneously breaks local gauge invariance, and can also be viewed as
forming a condensate which possesses a single-valued complex order parameter
with a well-defined local phase. From this viewpoint, the appearance of the
London penetration depth for a superconductor is analogous in an inverse
manner to the appearance of a mass for a gauge boson, such as that of the $W$
or $Z$ boson. \ Thus, the photon, viewed as a gauge boson, acquires a mass
inside the superconductor, such that its Compton wavelength becomes the
London penetration depth. Similar considerations apply to the effect of the
Chern-Simons term in the quantum Hall fluid. \ \qquad

Closely related to this spontaneous symmetry breaking process is the
appearance of Yang's off-diagonal long-range order (ODLRO) of the reduced
density matrix in the coordinate-space representation for most of these
macroscopically coherent quantum systems \cite{yang}. \ In particular, there
seems to be no limit on how far apart Cooper pairs can be inside a single
superconductor before they lose their quantum coherence. \ ODLRO and
spontaneous symmetry breaking are both purely quantum concepts with no
classical analogs. \ 

Within a quantum fluid, there should arise both the phenomenon of
instantaneous EPR correlations-at-a-distance, and the phenomenon of London's
``rigidity of the wavefunction,'' i.e., a Meissner-like response to
radiation fields. \ Both phenomena involve at the microscopic level $%
interactions$ of entangled particles with an external environment, either
through local $measurements$, such as in Bell-type measurements, or through
local $perturbations$, such as those arising from radiation fields
interacting locally with these particles. \ 

Although at first sight the notion of ``infinite quantum rigidity'' would
seem to imply infinite velocities, and hence would seem to violate
relativity, there are in fact no violations of relativistic causality here,
since the instantaneous EPR $correlations$-at-a-distance (as seen by an
observer in the center-of-mass frame) are not instantaneous $signals$%
-at-a-distance, which would instantaneously connect causes to effects \cite%
{ChiaoHeisenberg}. \ Also, experiments have verified the existence of
superluminal wave packet propagations, i.e., faster-than-$c$, infinite, and
even negative group velocities, for finite-bandwidth, analytic wave packets
in the excitations of a wide range of physical systems \cite{ChiaoSolvay}%
\cite{Superluminal}. \ An analytic function, e.g., a Gaussian wave packet,
contains sufficient information in its early tail such that a causal medium
can, during its propagation, reconstruct the entire wave packet with a
superluminal pulse advancement, and with little distortion. \ Relativistic
causality forbids\ only the $front$ velocity, i.e., the velocity of $%
discontinuities$ which connect causes to their effects, from exceeding the
speed of light $c$, but does not forbid a wave packet's $group$ velocity
from being superluminal. \ One example is the observed superluminal
tunneling of single-photon wave packets \cite{Steinberg}. \ Thus the notion
of ``infinite quantum rigidity,'' although counterintuitive, does not in
fact violate relativistic causality.

\section{The equivalence versus the uncertainty principle}

Concerning conceptual tension (2), the equivalence principle is formulated
at its outset using the concept of ``trajectory,'' or equivalently,
``geodesic.'' \ By contrast, Bohr has taught us that the very $concept$ of
trajectory must be abandoned at fundamental level, because of the
uncertainty principle. \ Thus the equivalence and the uncertainty principles
are in a fundamental conceptual tension. \ The equivalence principle is
based on the notion of locality, since it requires that the region of space,
inside which two trajectories of two nearby freely-falling objects of
different masses, compositions, or thermodynamic states, are to be compared,
go to zero volume, before the principle becomes exact. \ This limiting
procedure is in a conceptual tension with the uncertainty principle, since
taking the limit of the volume of space going to zero, within which these
objects are to be measured, makes their momenta infinitely uncertain. \
However, whenever the correspondence principle holds, the \textit{center of
mass} of a quantum wavepacket (for a single particle or for an entire
quantum object) moves according to Ehrenfest's theorem along a classical
trajectory, and $then$ it is possible to reconcile these two principles.

Davies \cite{Davies}\ has come up with a simple example of a quantum
violation of the equivalence principle \cite{onofrio}\cite{Adunas}\cite%
{Herdegen}: Consider two perfectly elastic balls, e.g., one made out of
rubber, and one made out of steel, bouncing against a perfectly elastic
table. \ If we drop the two balls from the same height above the table,
their classical trajectories, and hence their classical periods of
oscillation will be identical, and independent of the mass or composition of
the balls. \ This is a consequence of the equivalence principle. \ However,
quantum mechanically, there will be the phenomenon of tunneling, in which
the two balls can penetrate into the classically forbidden region $above$
their turning points. \ The extra time spent by the balls in the classically
forbidden region due to tunneling will depend on their mass (and thus on
their composition). \ Thus there will in principle be \textit{mass-dependent}
quantum corrections of the classical periods of the bouncing motion of these
balls, which will lead to quantum violations of the equivalence principle. \ 

There might exist macroscopic situations in which Ehrenfest's form of the
correspondence principle fails. \ Imagine that one is inside a macroscopic
quantum fluid, such as a big piece of superconconductor. \ Even in the limit
of a very large size and a very large number of particles inside this object
(i.e., in the thermodynamic limit), there exists no correspondence-principle
limit in which classical trajectories or geodesics for the \textit{relative
motion} of electrons which are members of Cooper pairs in Bohm singlet
states within the superconductor, make any sense. This is due to the
superposition principle and the entanglement of a macroscopic number of
identical particles inside these quantum fluids. Nevertheless, the \textit{%
motion of the center of mass} of the superconductor may obey perfectly the
equivalence principle, and may therefore be conceptualized in terms of a
geodesic.

\section{Quantum fluids as antennas for gravitational radiation\ }

Can the quantum rigidity arising from the energy gap of a quantum fluid
circumvent the problem of the tiny rigidity of classical matter, such as
that of the normal metals used in Weber bars, in their feeble responses to
gravitational radiation? ~One consequence of the tiny rigidity of classical
matter is the fact that the speed of sound in a Weber bar is typically five
orders of magnitude less than the speed of light. \ In order to transfer
energy coherently from a gravitational wave by classical means, for example,
by acoustical modes inside the bar\ to some local detector, e.g., a
piezoelectric crystal glued to the middle of the bar, the length scale of
the Weber bar $L$ is limited to a distance scale on the order of the speed
of sound times the period of the gravitational wave, i.e., an acoustical
wavelength $\lambda _{sound}$, which is typically five orders of magnitude
smaller than the gravitational radiation wavelength $\lambda $ to be
detected. \ This makes the Weber bar, which is thereby limited in its length
to $L\simeq \lambda _{sound}$, much too short an antenna to couple
efficiently to free space. \ 

However, rigid quantum objects, such as a two-dimensional electron gas in a
strong magnetic field which exhibits the quantum Hall effect, in what
Laughlin has called an ``incompressible quantum fluid'' \cite{Laughlin}, are
not limited by these classical considerations, but can have macroscopic
quantum phase coherence on a length scale $L$ on the same order as (or even
much greater than) the gravitational radiation wavelength $\lambda $. \
Since the radiation efficiency of a quadrupole antenna scales as the length
of the antenna $L$ to the fourth power when $L<<\lambda $, such quantum
antennas should be much more efficient in coupling to free space than
classical ones like the Weber bar by at least a factor of $\left( \lambda
/\lambda _{sound}\right) ^{4}$.

Weinberg\ gives a measure of the radiative coupling efficiency $\eta _{rad}$
of a Weber bar of mass $M$, length $L$, and velocity of sound $v_{sound}$,
in terms of a branching ratio for the emission of gravitational radiation by
the Weber bar, relative to the emission of heat, i.e., the ratio of the $%
rate $ of emission of gravitational radiation $\Gamma _{grav}$ relative to
the $rate$ of the decay of the acoustical oscillations into heat $\Gamma
_{heat}$, which is given by \cite{Weinberg}%
\begin{equation}
\eta _{rad}\equiv \frac{\Gamma _{grav}}{\Gamma _{heat}}=\frac{%
64GMv_{sound}^{4}}{15L^{2}c^{5}\Gamma _{heat}}\simeq {3\times 10^{-34}}%
\mbox{ ,}  
\label{Weinberg}
\end{equation}%
where $G$ is Newton's constant. The quartic power dependence of the
efficiency $\eta _{rad}$\ on the velocity of sound $v_{sound}$ arises from
the quartic dependence of the coupling efficiency to free space of a
quadrupole antenna upon its length $L$, when $L<<\lambda $. \ 

The long-range quantum phase coherence of a quantum fluid allows the typical
size $L$ of a quantum antenna to be comparable to the wavelength $\lambda $.
Thus the phase rigidity of the quantum fluid allows us in principle to
replace the velocity of sound $v_{sound}$ by the speed of light $c$.
Therefore, quantum fluids can be more efficient than Weber bars, based on
the $v_{sound}^{4}$ factor alone, by twenty orders of magnitude, i.e., 
\begin{equation}
\left( \frac{c}{v_{sound}}\right) ^{4}\simeq 10^{20}\mbox{ .}
\end{equation}%
Hence quantum fluids could be much more efficient receivers of this
radiation than Weber bars for detecting astrophysical sources of
gravitational radiation. This has previously been suggested to be the case
for superfluids and superconductors \cite{AnandanChiao}\cite{PengTorr}. \ 

Another important property of quantum fluids lies in the fact that they can
possess an extremely low dissipation coefficient $\Gamma _{heat}$, as can be
inferred, for example, by the existence of persistent currents in
superfluids that can last for indefinitely long periods of time. Thus the
impedance matching of the quantum antenna to free space \cite{Impedance}, or
equivalently, the branching ratio of energy emitted into the gravitational
radiation channel rather than into the heat channel, can be much larger than
that calculated above for the classical Weber bar.\qquad

\section{Minimal-coupling rule for a quantum Hall fluid}

The electron, which possesses charge $e$, rest mass $m$, and spin $s=1/2$,
obeys the Dirac equation. The nonrelativistic, interacting, fermionic
many-body system, such as that in the quantum Hall fluid, should obey the
minimal-coupling rule which originates from the covariant-derivative
coupling of the Dirac electron to curved spacetime, viz., \cite{DaviesBook}%
\cite{Weinberg} 
\begin{equation}
p_{\mu }\rightarrow p_{\mu }-eA_{\mu }-\frac{1}{2}\Sigma _{AB}\omega _{\mu
}^{AB}
\end{equation}%
where $p_{\mu }$ is the electron's four-momentum, $A_{\mu }$ is the
electromagnetic four-potential, $\Sigma _{AB}$ are the Dirac $\gamma $
matrices in curved spacetime with tetrad (or vierbein) $A,B$ indices, and $%
\omega _{\mu }^{AB}$ are the components of the spin connection%
\begin{equation}
\omega _{\mu }^{AB}=e^{A\nu }\nabla _{\mu }\left. e^{B}\right. _{\nu }
\end{equation}%
where $e^{A\nu }$ and $\left. e^{B}\right. _{\nu }$ are tetrad four-vectors,
which are sets of four orthogonal unit vectors of spacetime, such as those
corresponding to a local inertial frame.

Spacetime curvature directly affects the phase of the wavefunction, 
leading to fringe shifts of quantum-mechanical interference patterns within 
atomic interferometers \cite{GRGessay}. Moreover, it is well known that 
the vector potential $A_{\mu }$ will also lead to a quantum interference effect, in 
which the gauge-invariant Aharonov-Bohm phase becomes observable. Similarly, 
the spin connection $\omega _{\mu }^{AB}$, in its Abelian holonomy, should 
also lead to a quantum interference effect, in which the gauge-invariant 
Berry phase \cite{Chiao-Wu} becomes observable. The following Berry phase 
picture of a spin coupled to curved spacetime leads to an intuitive way of 
understanding why there could exist a coupling between a classical GR wave 
and a classical EM wave mediated by a quantum fluid with charge and spin, such  
as the quantum Hall fluid.

Due to its gyroscopic nature, the spin vector of an electron undergoes \emph{%
parallel transport} during the passage of a GR wave. The spin of the
electron is constrained to lie inside the space-like submanifold of curved
spacetime. This is due to the fact that we can always transform to a
co-moving frame, such that the electron is at rest at the origin of this
frame. In this frame, the spin of the electron must be purely a space-like
vector with no time-like component. This imposes an important $constraint$
on the motion of the electron's spin, such that whenever the space-like
submanifold of spacetime is disturbed by the passage of a gravitational
wave, the spin must remain at all times $perpendicular$ to the local time
axis. If the spin vector is constrained to follow a conical trajectory
during the passage of the gravitational wave, the electron picks up a Berry
phase proportional to the solid angle subtended by this conical trajectory
after one period of the GR wave.

In a manner similar to the persistent currents induced by the Berry phase in
systems with ODLRO \cite{Lyanda-Geller}, such a Berry phase induces an
electrical current in the quantum Hall fluid, which is in a macroscopically
coherent ground state \cite{Girvin}. This macroscopic current generates an
EM wave. Thus a GR wave can be converted into an EM wave. By reciprocity,
the time-reversed process of the conversion from an EM wave to a GR wave
must also be possible.

In the nonrelativistic limit, the four-component Dirac spinor is reduced to
a two-component spinor. While the precise form of the nonrelativistic
Hamiltonian is not known for the many-body system in a weakly curved
spacetime consisting of electrons in a strong magnetic field, I conjecture
that it will have the form%
\begin{equation}
H=\frac{1}{2m}\left( p_{i}-eA_{i}-\frac{1}{2}\sigma_{ab} {\Omega}%
_{i}^{ab}\right) ^{2}+V
\end{equation}%
where $i$ is a spatial index, $a,b$ are spatial tetrad incides, $\sigma_{ab}$
is a two-by-two matrix-valued tensor representing the spin \cite{spin}, and $%
\sigma_{ab}{\Omega}_{i}^{ab}$ is the nonrelativistic form of $\Sigma
_{AB}\omega _{\mu }^{AB}$. Here $H$ and $V$ are two-by-two matrix operators
on the two-component spinor electron wavefunction in the nonrelativistic
limit. The potential energy $V$ includes the Coulomb interactions between
the electrons in the quantum Hall fluid. This nonrelativistic Hamiltonian
has the form%
\begin{equation}
H=\frac{1}{2m}\left( \mathbf{p}-\mathbf{a}-\mathbf{b}\right) ^{2}+V\mbox{ ,}
\end{equation}%
where the particle index, the spin, and the tetrad indices have all been
suppressed. Upon expanding the square, it follows that for a quantum Hall
fluid of uniform density, there exists a cross-coupling or interaction
Hamiltonian term of the form%
\begin{equation}
H_{int}\sim \mathbf{a}\cdot \mathbf{b}\mbox{ ,}  \label{cross-coupling}
\end{equation}%
which couples the electromagnetic $\mathbf{a}$ field to the gravitational $%
\mathbf{b}$~field. In the case of time-varying fields, $\mathbf{a}(t)$ and $%
\mathbf{b}(t)$ represent EM and GR radiation, respectively.

In first-order perturbation theory, the quantum adiabatic theorem predicts
that there will arise the cross-coupling energy between the two radiation
fields mediated by the quantum fluid 
\begin{equation}
\Delta E\sim \langle \Psi _{0}|\mathbf{a}\cdot \mathbf{b}|\Psi _{0}\rangle 
\end{equation}%
where $|\Psi _{0}\rangle $ is the unperturbed ground state of the system.
For the adiabatic theorem to hold, there must exist an energy gap $E_{gap}$
(e.g., the quantum Hall energy gap) separating the ground state from all
excited states, in conjunction with the approximation that the time
variation of the radiation fields must be slow compared to the gap time $%
\hbar /E_{gap}$. This suggests that under these conditions, there might
exist an interconversion process between these two kinds of classical
radiation fields mediated by this quantum fluid, as indicated in Figure 4.

The question immediately arises: EM radiation is fundamentally a spin 1
(photon) field, but GR radiation is fundamentally a spin 2 (graviton) field.
How is it possible to convert one kind of radiation into the other, and not
violate the conservation of angular momentum? ~The answer: The EM wave
converts to the GR wave \emph{through a medium}. Here specifically, the
medium of conversion consists of a strong DC magnetic field applied to a
system of electrons. This system possesses an axis of symmetry pointing
along the magnetic field direction, and therefore transforms like a spin 1
object. When coupled to a spin 1 (circularly polarized) EM radiation field,
the total system can in principle produce a spin 2 (circularly polarized) GR
radiation field, by the addition of angular momentum. However, it remains an
open question as to how strong this interconversion process is between EM
and GR radiation. Most importantly, the size of the conversion efficiency of
this transduction process needs to be determined by experiment.

We can see more clearly the physical significance of the interaction
Hamiltonian $H_{int}\sim \mathbf{a}\cdot \mathbf{b}$ once we convert it into
second quantized form and express it in terms of the creation and
annihilation operators for the positive frequency parts of the\ two kinds of
radiation fields, as in the theory of quantum optics, so that in the
rotating-wave approximation%
\begin{equation}
H_{int}\sim a^{\dagger }b+b^{\dagger }a\text{ ,}
\end{equation}%
where the annihilation operator $a$ and the creation operator $a^{\dagger }$
of the single classical mode of the plane-wave EM radiation field
corresponding the $\mathbf{a}$ term, obey the commutation relation $%
[a,a^{\dagger }]=1$, and where the annihilation operator $b$ and the
creation operator $b^{\dagger }$ of the single classical mode of the
plane-wave GR radiation field corresponding to the $\mathbf{b}$ term, obey
the commutation relation $[b,b^{\dagger }]=1$. \ (This represents a crude,
first attempt at quantizing the gravitational field, which applies only in
the case of weak, linearized gravity.) \ The first term $a^{\dagger }b$ then
corresponds to the process in which a graviton is annihilated and a photon
is created inside the quantum fluid, and similarly the second term $%
b^{\dagger }a$ corresponds to the reciprocal process, in which a photon is
annihilated and a graviton is created inside the quantum fluid.

Let us return once again to the question of whether there exists $any$
difference in the response of quantum fluids to tidal fields in
gravitational radiation, and the response of classical matter, such as the
lattice of ions in a superconductor, for example, to such fields. The
essential difference between quantum fluids and classical matter is the
presence or absence of macroscopic quantum phase coherence. In quantum
matter, there exist quantum interference effects, whereas in classical
matter, such as in the lattice of ions of a superconductor, decoherence
arising from the environment destroys any such interference. As argued
earlier in section 3, the response of quantum fluids and of classical matter
to these fields will therefore differ from each other.

In the case of superconductors, Cooper pairs of electrons possess a
macroscopic phase coherence, which can lead to an Aharonov-Bohm-type
interference absent in the ionic lattice. Similarly, in the quantum Hall
fluid, the electrons will also possess macroscopic phase coherence \cite%
{Girvin}, which can lead to Berry-phase-type interference absent in the
lattice. Furthermore, there exist ferromagnetic superfluids with intrinsic
spin, in which an ionic lattice is completely absent, such as in
spin-polarized atomic BECs \cite{CornellWieman} 
and in superfluid helium 3 \cite{Osheroff}. In
such ferromagnetic quantum fluids, there exists no ionic lattice to give
rise to any classical response which could prevent a quantum response to
tidal gravitational radiation fields. The Berry-phase-induced response of
the ferromagnetic superfluid arises from the spin connection (see the above
minimal-coupling rule, which can be generalized from an electron spin to a
nuclear spin coupled to the curved spacetime associated with gravitational
radiation), and leads to a purely quantum response to this radiation. The
Berry phase induces time-varying macroscopic quantum flows in this
ferromagnetic ODLRO system \cite{Lyanda-Geller}, which transports
time-varying orientations of the nuclear magnetic moments. This
ferromagnetic superfluid can therefore also in principle convert
gravitational into electromagnetic radiation, and vice versa, in a manner
similar to the case discussed above for the ferromagnetic quantum Hall fluid.

Thus we expect there to exist differences between classical and quantum
fluids in their respective linear responses to weak external perturbations
associated with gravitational radiation. Like superfluids, the quantum Hall
fluid is an example of a quantum fluid which differs from a classical fluid
in its current-current correlation function \cite{Forster} in the presence
of GR waves. In particular, GR waves can induce a transition of the quantum
Hall fluid out of its ground state $only$ by exciting a quantized,
collective excitation across the quantum Hall energy gap. This collective
excitation would involve the correlated motions of a macroscopic number of
electrons in this coherent quantum system. Hence the quantum Hall fluid is
effectively incompressible and dissipationless, and is thus a good candidate
for a quantum antenna.

There exist other situations in which a minimal-coupling rule similar to the
one above, arises for $scalar$ quantum fields in curved spacetime. DeWitt %
\cite{DeWitt} suggested~in 1966 such a coupling in the case of
superconductors \cite{Solli}. Speliotopoulos \cite{Speliotopoulos1995} noted
in 1995 that a cross-coupling term of the form $H_{int}\sim \mathbf{a}\cdot 
\mathbf{b}$ arose in the long-wavelength limit of a certain quantum
Hamiltonian derived from the geodesic deviation equations of motion using
the transverse-traceless gauge for GR waves.

Speliotopoulos and I have been working on the problem of the coupling of a
scalar quantum field to curved spacetime in a general laboratory frame,
which avoids the use of the long-wavelength approximation \cite{GLF}. In
general relativity, there exists in general no global time coordinate that
can apply throughout a large system, since for nonstationary metrics, such
as those associated with gravitational radiation, the local time axis varies
from place to place in the system. It is therefore necessary to set up
operationally a general laboratory frame by which an observer can measure
the motion of slowly moving test particles in the presence of weak,
time-varying gravitational radiation fields.

For either a classical or quantum test particle, the result is that its mass 
$m$ should enter into the Hamiltonian through the replacement of $\mathbf{p}%
-e\mathbf{A}$ by $\mathbf{p}-e\mathbf{A}-m\mathbf{N}$, where $\mathbf{N}$ is
the small, local tidal velocity field induced by gravitational radiation on
a test particle located at $X_a$ relative to the observer at the origin
(i.e., the center of mass) of this frame, where, for the small deviations $%
h_{ab}$ of the metric from that of flat spacetime, 
\begin{equation}
N_a=\frac{1}{2}\int_{0}^{X_a} \frac{\partial h_{ab}}{\partial t} dX^{b}.
\end{equation}
Due to the quadrupolar nature of gravitational tidal fields, the velocity
field $\mathbf{N}$ for a plane wave grows linearly in magnitude with the
distance of the test particle from the center of mass, as seen by the
observer located at the center of mass of the system. Therefore, in order to
recover the standard result of classical GR that only $tidal$ gravitational
fields enter into the coupling of radiation and matter, one expects in
general that a new characteristic length scale $L$ corresponding to the
typical size of the distance $X_a$ separating the test particle from the
observer, must enter into the determination of the coupling constant between
radiation and matter. For example, $L$ can be the typical size of the
detection apparatus (e.g., the length of the arms of the Michelson
interferometer used in LIGO), or of the transverse Gaussian wave packet size
of the gravitational radiation, so that the coupling constant associated
with the Feynman vertex for a graviton-particle interaction becomes
proportional to the $extensive$ quantity $\sqrt{G}L$, instead of an $%
intensive$ quantity involving only $\sqrt{G}$.

For the case of superconductors, treating Cooper pairs of electrons as
bosons, we would expect the above arguments would carry over with the charge 
$e$ replaced by $2e$ and the mass $m$ replaced by $2m$. For quantum fluids
which possess an order parameter $\Psi $ obeying the Ginzburg-Landau
equation, the above minimal-coupling rule suggests that this equation be
generalized as follows:%
\begin{equation}
\frac{1}{2m}\left( \frac{\hbar }{i}\mathbf{\nabla }-\mathbf{a}-\mathbf{b}%
\right) ^{2}\Psi +\beta |\Psi |^{2}\Psi =-\alpha \Psi \text{ ,}
\label{Generalized GL equation}
\end{equation}
where $\mathbf{b} \sim\mathbf{N}$.

\section{Quantum transducers between EM and GR waves?\ \qquad\ \ \ }

Returning to the general problem of quantum fields embedded in curved
spacetime, we recall that the ground state of a superconductor, which
possesses spontaneous symmetry breaking, and therefore ODLRO, is very
similar to that of the physical vacuum, which is believed also to possess
spontanous symmetry breaking through the Higgs mechanism. In this sense,
therefore, the vacuum is ``superconducting.'' The question thus arises: How
does a ground or ``vacuum'' state of a superconductor, and other quantum
fluids viewed as ground states of nonrelativistic quantum field theories
with ODLRO, interact with dynamically changing spacetimes, e.g., a GR
wave? We believe that this question needs both theoretical and experimental
investigation.

In particular, motivated by the discussion in the previous section, we
suspect that there might exist superconductors, viewed as quantum fluids,
which are transducers between EM and GR waves based on the cross-coupling
Hamiltonian $H_{int}$ $\sim \mathbf{a}\cdot \mathbf{b}$. One possible
geometry for an experiment is shown in Figure 4. An EM\ wave impinges on the
quantum fluid, which converts it into a GR wave in process (a). In the
time-reversed process (b), a GR wave impinges on the quantum fluid, which
converts~it back into an EM wave. It is an open question at this point as to
what the conversion efficiency of such quantum transducers will be \cite%
{enhancement}. This question is best settled by an experiment to measure
this efficiency by means of a Hertz-type apparatus, in which process (a) is
used for generating gravitational radiation, and process (b), inside a
separate quantum transducer, is used to detect this radiation.

If the quantum transducer conversion efficiency turns out to be high, this
will lead to an avenue of research which could be called ``gravity radio.''
~I have performed a preliminary version of this Hertz-type experiment with
Walt Fitelson using the high $T_{c}$ superconductor YBCO to measure its
transducer efficiency at microwave frequencies. We have obtained an upper
limit on the conversion efficiency for YBCO at liquid nitrogen temperature
of $1.6\times 10^{-5}$. Details of this experiment will be reported
elsewhere \cite{Venice}.

\section{Conclusions}

The conceptual tensions between QM and GR, the two main fields of interest
of John Archibald Wheeler, could indeed lead to important experimental
consequences, much like the conceptual tensions of the past. \ I have
covered here in detail only one of these conceptual tensions, namely, the
tension between the concept of \textit{spatial nonseparability} of physical
systems due to the notion of nonlocality embedded in the superposition
principle, in particular, in the entangled states of QM, and the concept of 
\textit{spatial separability} of all physical systems due to the notion of
locality embedded in the equivalence principle in GR. \ This has led to the
idea of antennas and transducers using quantum fluids as potentially
practical devices, which could possibly open up a door for further exciting
discoveries \cite{CMB}.

\section{Acknowledgments}

I dedicate this paper to my teacher, John Archibald Wheeler, whose vision
helped inspire this paper. I am grateful to the John Templeton Foundation
for the invitation to contribute to this Volume, and would like to thank my
father-in-law, the late Yi-Fan Chiao, for his financial and moral support of
this work. \ This work was supported also by the ONR.

\pagebreak

\section{Figure Captions}

\begin{enumerate}
\item Figure 1: \textit{Three intersecting circles} in a Venn-like diagram
represent the three main pillars of physics at the beginning of the 21st
century.\ The top circle represents quantum mechanics, and is labeled by
Planck's constant $\hbar $. The left circle represents relativity, and is
labeled by the two constants $c$, the speed of light, and $G$, Newton's
constant.\ The right circle represents statistical mechanics and
thermodynamics, and is labeled by Boltzmann's constant $k_{B}$. Conceptual
tensions exist at the intersections of these three circles, which may lead
to fruitful experimental consequences.

\item Figure 2: \textit{Liquid drop} placed at the center of a not-to-scale
sketch of the Space Station, where it is subjected to the tidal force due to
the Earth's gravity. Is there any difference between the shape of a
classical and a quantum liquid drop, for example, between a drop of water
and one composed of superfluid helium?

\item Figure 3: \textit{Whispering gallery modes} of a liquid drop arise in
the correspondence principle limit, when an atom or a photon wave packet
bounces at grazing incidence off the inner surface of the drop in multiple
specular internal reflections, to form a closed polygonal trajectory. The
Bohr-Sommerfeld quantization rule leads to a discrete set of such modes.

\item Figure 4: \textit{Quantum transducer} between electromagnetic (EM) and
gravitational (GR) radiation, consisting of a quantum fluid with charge and
spin, such as the quantum Hall fluid. The minimal-coupling rule for an
electron coupled to curved spacetime via its charge and spin, results in two
processes. In process (a) an EM plane wave is converted upon reflection from
the quantum fluid into a GR plane wave; in process (b), which is the
reciprocal or time-reversed process, a GR plane wave is converted upon
reflection from the quantum fluid into an EM plane wave. Transducer
interconversion between these two kinds of waves may also occur upon $%
transmission$ through the quantum fluid, as well as upon $reflection$.
\end{enumerate}

\pagebreak

\begin{figure}[tbp]
\centerline{\includegraphics{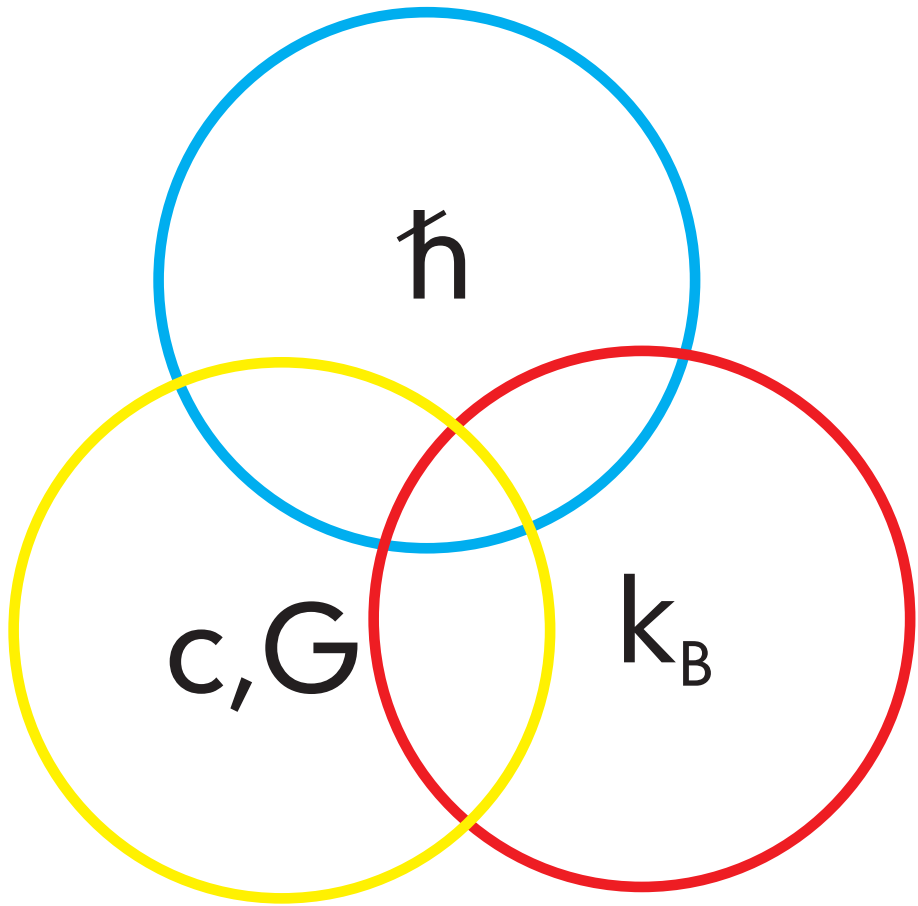}}
\caption{Three intersecting circles...}
\end{figure}

\begin{figure}[tbp]
\centerline{\includegraphics{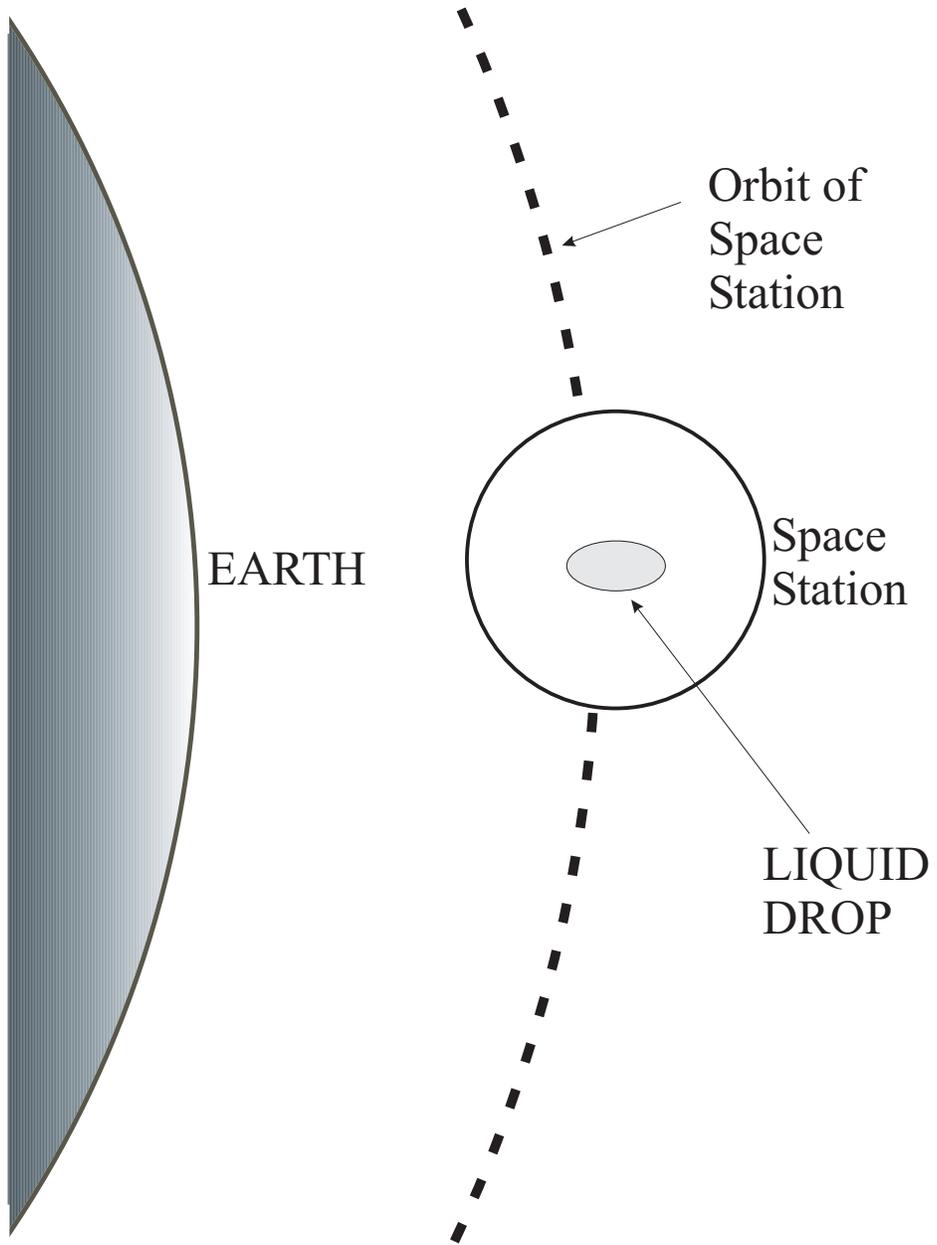}}
\caption{Liquid drop...}
\end{figure}

\begin{figure}[tbp]
\centerline{\includegraphics{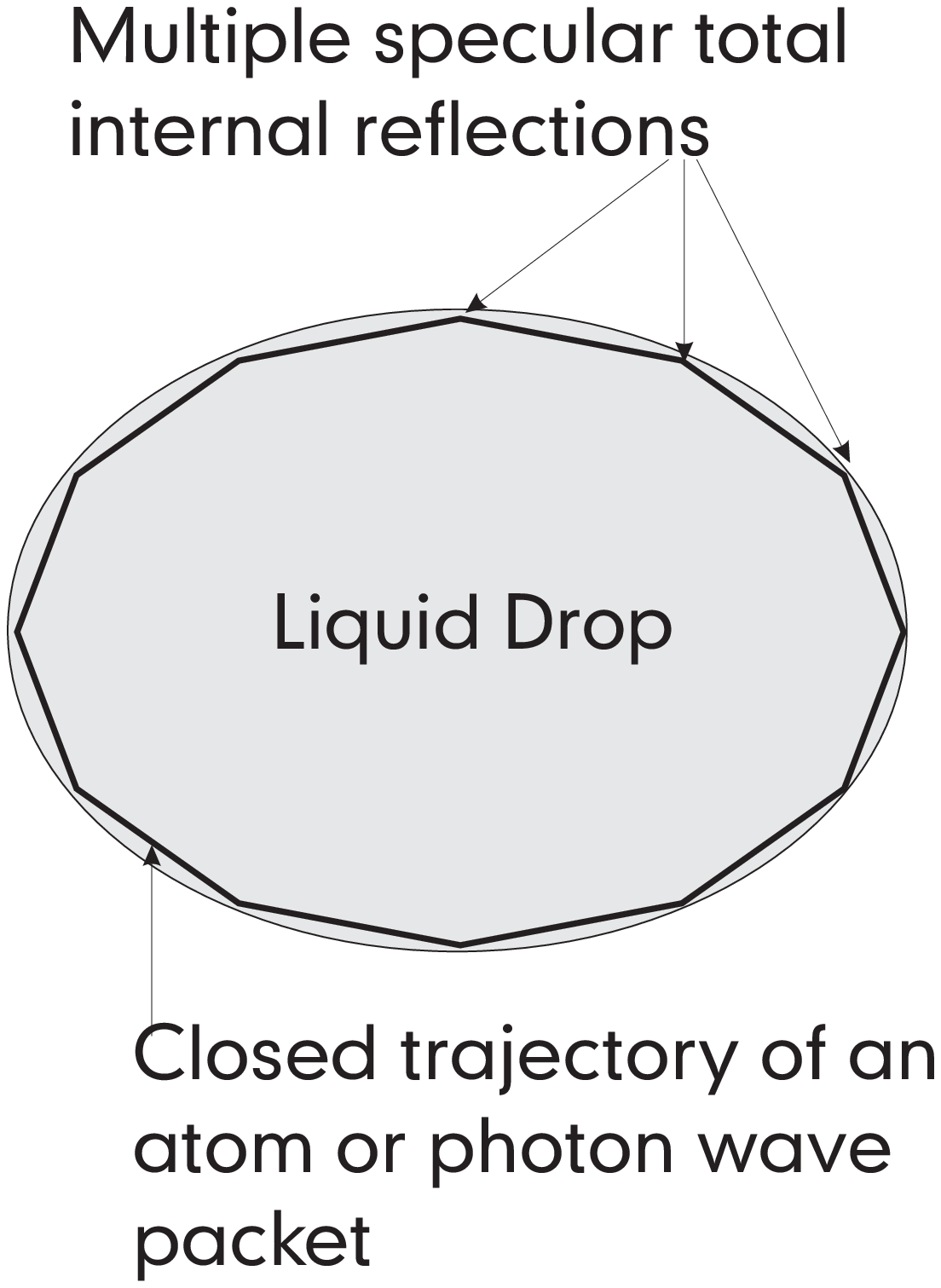}}
\caption{Whispering gallery modes...}
\end{figure}

\begin{figure}[tbp]
\centerline{\includegraphics{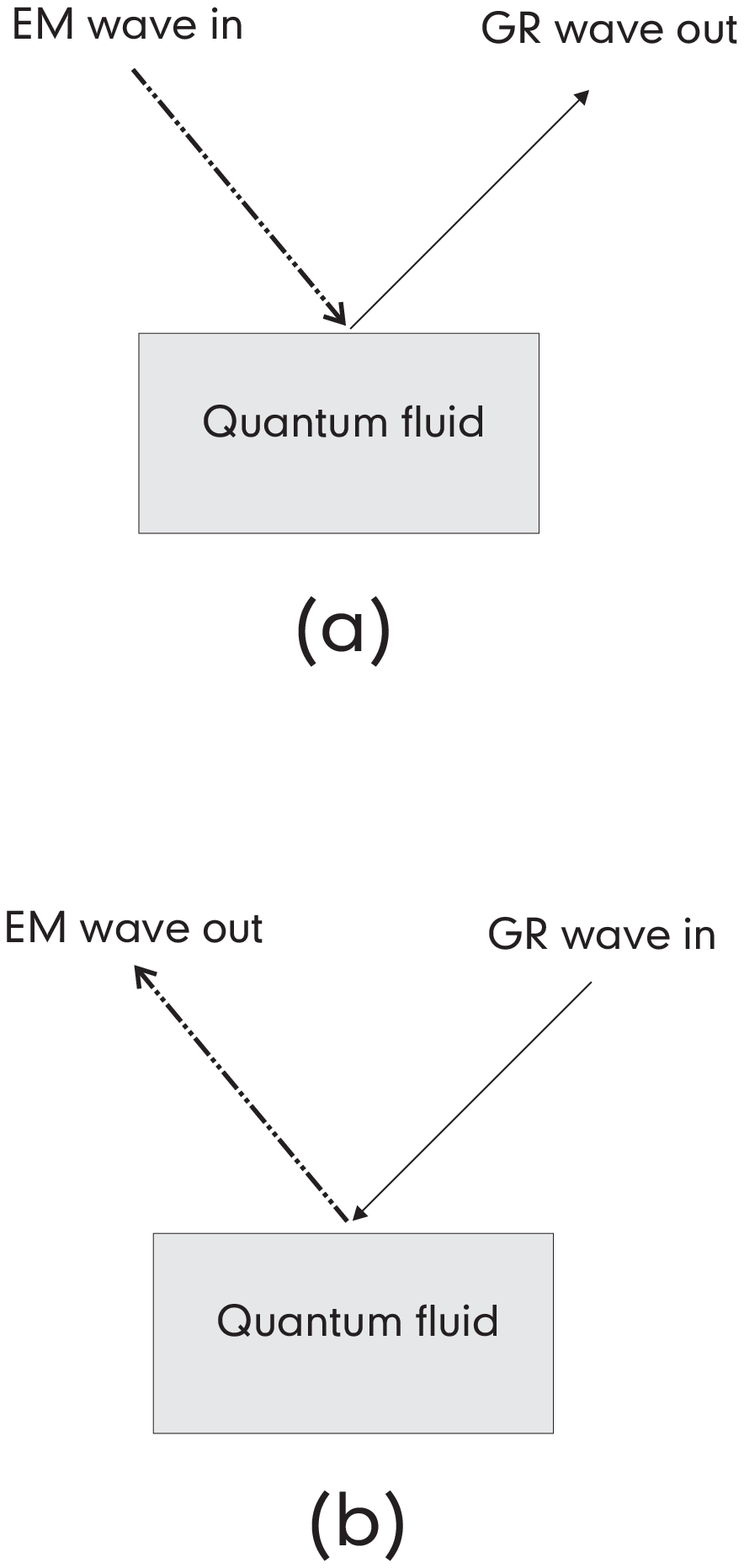}}
\caption{Quantum transducer...}
\end{figure}


\begin{thebibliography}{99}
\bibitem{enhancement} One might expect a transducer power conversion
efficiency of $Gm^{2}\cdot 4\pi \varepsilon _{0}/e^{2}=2.4\times 10^{-43}$
based on a naive classical picture in which each individual electron follows
a deterministic, Newtonian trajectory. If this classical picture had been
correct, there would have been no hope of actually observing this conversion
process, based on the limited sensitivity of existing experimental
techniques such as those described in \cite{Venice}. However,
superconductivity is fundamentally a quantum mechanical phenomenon. Due to
the macroscopic coherence of the ground state with ODLRO, and the existence
of a non-zero energy gap, there may exist quantum many-body enhancements to
this classical conversion efficiency. In addition to these enhancements,
there must exist additional enhancements due to the fact that the intensive
coupling constant $\sqrt{G}$ of the Feynman graviton-matter vertex should be
replaced by the extensive coupling constant $\sqrt{G}L$, in order to account
correctly for the $tidal$ nature of GR waves \cite{GLF}. Furthermore, it is difficult 
to calculate the branching ratio for converting EM waves into GR waves versus into heat, since it is 
difficult to predict theoretically the rate of dissipation into heat $\Gamma_{heat}$ in 
Eq.(\ref{Weinberg}) for quantum fluids. Hence an experiment is needed to measure 
this conversion efficiency.
\bibitem{CMB} Quantum transducers, if sufficiently efficient, would allow us
to directly observe for the first time the CMB (Cosmic Microwave Background)
in GR radiation, which would tell us much about the very early Universe.
\end{thebibliography}
\end{document}